\documentclass[showpacs,amsmath,amssymb,prl,superscriptaddress,floatfix,twocolumn]{revtex4}
\usepackage{graphicx}
\usepackage{dcolumn}
\usepackage{bm}

\begin{document}

\preprint{UA-NPPS/01/2005}
\title{Hyperacceleration in a stochastic Fermi-Ulam model}
\author{A.K. Karlis}
\author{P.K. Papachristou}
\author{F.K. Diakonos}
\email[]{fdiakono@phys.uoa.gr}
\affiliation{Department of Physics, University of Athens, GR-15771 Athens, Greece}
\author{V. Constantoudis}
\affiliation{Institute of Microelectronics, NCSR Demokritos, P.O.
Box 60228, Attiki, Greece}
\author{P. Schmelcher}
\affiliation{Physikalisches Institut,  Universit\"at Heidelberg,
Philosophenweg 12, 69120 Heidelberg, Germany}
\affiliation{Theoretische Chemie, Im Neuenheimer Feld 229,
Universit\"at Heidelberg, 69120 Heidelberg, Germany}
\date{\today}
\begin{abstract}
Fermi acceleration in a Fermi-Ulam model, consisting of an ensemble
of particles bouncing between two, infinitely heavy, stochastically
oscillating hard walls, is investigated. It is shown that the widely
used approximation, neglecting the displacement of the walls (static
wall approximation), leads to a systematic underestimation of
particle acceleration. An improved approximative map is introduced,
which takes into account the effect of the wall displacement, and in
addition allows the analytical estimation of the long term behavior
of the particle mean velocity as well as the corresponding
probability distribution, in complete agreement with the numerical
results of the exact dynamics. This effect accounting for the
increased particle acceleration --{\it Fermi hyperacceleration}-- is
also present in higher dimensional systems, such as the driven
Lorentz gas.
\end{abstract}
\maketitle

In 1949 Fermi \cite{Fermi:1949} proposed an  acceleration mechanism
of cosmic ray particles interacting with a time dependent magnetic
field (for a review see \cite{Blandford:1987}). Ever since, this has
been a subject of intense study in a broad range of systems in
various areas of physics, including astrophysics
\cite{Veltri:2004,Kobayakawa:2002,Malkov:1998}, plasma physics
\cite{Michalek:1999,Milovanov:2001}, atom optics
\cite{Saif:1998,Steane:1995} and has even been used for the
interpretation of experimental results in atomic physics
\cite{Lanzano:1999}. Furthermore, when the mechanism is linked to
higher dimensional time-dependent billiards, such as a
time-dependent variant of the classic Lorentz Gas, it has profound
implications on statistical and solid state physics \cite{Loskutov}.
Several modifications of the original model have been suggested, one
of which is the well-known Fermi-Ulam model (FUM)
\cite{Ulam:1961,Lieberman:1972,Lichtenberg:1992} which describes the
bouncing of a ball between an oscillating and a fixed wall. FUM and
its variants have been the subject of extensive theoretical (see
Ref.~\cite{Lieberman:1972} and references therein) and experimental
\cite{Kowalik:1988,Warr:1996,Celaschi:1987} studies as they are
simple to conceive but hard to understand in that their behavior is
quite complicated. A standard simplification \cite{Lieberman:1972}
widely used in the literature, the static wall approximation (SWA),
ignores the displacement of the moving wall but retains the time
dependence in the momentum exchange between particle and wall at the
instant of collision as if the wall were oscillating. The SWA speeds
up time-consuming numerical simulations and allows semi-analytical
treatments as well as a deeper understanding of the system
\cite{Lieberman:1972,Lichtenberg:1980,Leonel:2004a,Leonel:2004b,Leonel:2005}.
However, as shown by Einstein in his treatment of the Brownian
random walk \cite{Einstein:1956}, taking account of the full phase
space trajectory (instead of the momentum component only) is
essential for the correct description of diffusion processes. More
recently, in the context of diffusion in the deterministic FUM,
Lieberman \emph{et al} have shown that one has to employ both
canonical conjugate variables (position and momentum) in order to
obtain the correct momentum distribution in the asymptotic steady
state \cite{Lichtenberg:1980}. The present work shows that even in
the absence of an asymptotic steady state the diffusion in velocity
space is deeply affected by the location of the collision events in
configuration space.

The dynamical system in question consists of two harmonically driven
infinitely heavy walls with an ensemble of particles bouncing
between them. When a particle collides with a certain wall, a random
shift of the phase of the other wall, which is uniformly distributed
in $[0,2 \pi)$, occurs. The stochastic component in the oscillation
law of the wall simulates the influence of a thermal environment on
wall motion and leads to Fermi acceleration
\cite{Zaslavskii:1965,Brahic:1971,Lieberman:1972,Lichtenberg:1980,Leonel:2004a,Leonel:2005}.
It should be noted that although stochasticity can be introduced
otherwise --for instance, via a random component in the angular
frequency of oscillation-- the random phase approach has become
quite common as a method of randomization of the FUM and its
modifications \cite{Lieberman:1972,Loskutov,Leonel:2005}, partially
because it is the only conceivable way to randomize the system
without changing the energy of the moving wall.

Despite the external randomization \cite{Lieberman:1972} the SWA
does not provide an accurate description of the diffusion process;
more specifically, the energy gain of the particle is substantially
underestimated. For this reason we introduce in this Letter the
so-called hopping wall approximation (HWA), which takes into account
the effect of the wall displacement. By means of this approximation
it is made clear how the oscillation of the wall in configuration
space affects the acceleration law of an ensemble of particles.
Furthermore, the corresponding map allows analytical treatment and
is as computationally efficient as the SWA and it enables us to
calculate the evolution of the velocity distribution of the
particles for long time periods.

The specific setup of the studied system is determined by the
oscillation frequencies $\omega_i$ and amplitudes $A_i$ of the two
walls ($i=L, R$) as well as the distance $w$ between the walls at
equilibrium. However, the dynamics does not depend on each of these
parameters explicitly. It is therefore appropriate to introduce the
relevant dimensionless quantities $\epsilon_i=\frac{A_i}{w}$,
$r=\frac{\omega_L}{\omega_R}$. Obviously, when the ratio
$\chi=\frac{\epsilon_L}{\epsilon_R}$ meets the condition $\chi\gg 1$
(or $\chi\ll1$) the original FUM is recovered. For the sake of
simplicity the case $\chi=1$ ($A_L=A_R=A$) and $r=1$
($\omega_L=\omega_R=\omega$) is exclusively considered in the
following. Using as a length unit the spacing between the walls $w$
and as a time unit $\frac{1}{\omega}$  the dynamical laws of the
system can be derived in a dimensionless form:
\begin{equation}
V_n=-V_{n-1} + 2 u_n ~~~;~~~u_n = \epsilon\cos(\delta t_n + t_{n-1}
+ \eta_n) \label{eq:eq1}
\end{equation}
where, $V_n$ denotes particle velocity after the $n$th collision,
$u_n$ for wall velocity on collision and $\eta_n$ the random phase
component.  The time of free  flight $\delta t_n$ is obtained by
solving the implicit equation
\begin{equation}
X_{n-1} + V_{n-1} \delta t_n  = \pm \frac{1}{2} + \epsilon
\sin(\delta t_n + t_{n-1} + \eta_n) \label{eq:eq2}
\end{equation}
 Obviously, Eq.~(\ref{eq:eq2}) links
the time of the $n$th collision to the position $X_{n-1}$ of the
particle in the previous collision. The SWA simplifies the process
on the basis of the assumption that the time of free flight, $\delta
t_n$, depends only on particle velocity, i.e. $\delta t_n=
\frac{\pm1}{V_{n-1}}$. If this approximation is applied to the
system of two oscillating walls, then it is possible to extract
analytically the ensemble averaged velocity square $\langle \delta
V_n^2 \rangle=\langle V_n^2 - V_{n-1}^2 \rangle$ of the particle
after integration over the random phase $\eta_n$: $\langle \delta
V^2_n \rangle=2\epsilon^2$. Given that $\langle V_n^2 \rangle =
\sum\limits_{i = 1}^n \langle \delta V_i^2 \rangle + \langle V_0^2
\rangle$, the root mean square particle velocity is:
\begin{equation}
V_{rms,~SWA}(n) = \sqrt{2\epsilon^2\cdot n + {\langle V_0}^2\rangle}
\label{eq:eq3}
\end{equation}

Additionally, this quantity can be determined numerically using the
exact dynamical law given in eqs.(\ref{eq:eq1}),(\ref{eq:eq2}). The
calculations are performed on the basis of an ensemble of $10^4$
trajectories with $\langle V_0\rangle=\frac{10^2}{15}$,
$\epsilon=\frac{1}{15}$. Corresponding results are presented in
Fig.~1 together with the analytical result (\ref{eq:eq3}), and show
that there is a considerable difference between the acceleration
rate of the root mean square (RMS) velocity given by the exact map
and by the SWA. For a  better understanding of how this difference
originates, we improve the SWA by taking into account the impact of
the displacement of the walls incorporated in the exact dynamics. As
particle velocity increases the time of free flight decreases,
making it possible to approximate the position of the scatterer at
the instant of the $n$th collision with that at the $(n-1)$th
collision. This approximation allows for an analytical evaluation of
$\delta t_n$ and defines the hopping wall approximation i.e. HWA. In
this framework the time interval $\delta t_n$ reads as follows:
\begin{equation}
\delta {t_n} = \delta t_n^\ast \pm \frac{1}{V_{n-1}} \label{eq:eq4}
\end{equation}
where, $\delta t_n^{\ast} = \frac{\epsilon \left[\sin({t}_{n - 1} +
\eta _{n}) - \sin({t}_{n - 2} + \eta _{n-1})\right] }{{V}_{n-1}}$ is
the correction term to the time of free flight predicted by SWA.

In order to derive $\langle \delta V_n^2 \rangle $ using
eqs.~(\ref{eq:eq1}), (\ref{eq:eq4}) the following integrals have to
be calculated:
\begin{equation}
 I_j = \int_0^{2\pi } \int_0^{2\pi}  \frac{1}{4\pi
^2} \bigg[\epsilon\cos \Big(t_{n - 1}+\delta{t_n}+\eta _n
\Big)\bigg]^j d \eta_n d \eta_{n - 1} \label{eq:eq6}
\end{equation}
where $j=1,2$. An exact analytical calculation of these integrals is
not possible. However, for the set of parameters considered here,
$\delta t_n^{\ast}$ is much smaller compared to the other phase
components. Therefore, we expand the r.h.s. of eq.~(\ref{eq:eq6}) to
the leading order of $\delta t_n^{\ast}$ and then integrate over
$\eta_n$ and $\eta_{n - 1}$ which yields:
\begin{equation}
I_1 \approx - \frac{\epsilon^2\cos \left(\frac{1}{\langle V_{n -
1}\rangle }\right)}{2 \langle V_{n - 1}\rangle }
\nonumber~~~~~;~~~~~ I_2 \approx \frac{\epsilon^2}{2} \label{eq:eq7}
\end{equation}
Therefore, we find:
\begin{equation}
\langle \delta {V}_n^2 \rangle \approx 2
\epsilon^2\cos\left(\frac{1}{\epsilon\langle {V}_{n - 1}
\rangle}\right) + 2\epsilon^2 \label{eq:eq8}
\end{equation}
In the limit of high particle velocities  $ \langle V_{n - 1}\rangle
\gg 1 $, eq.~(\ref{eq:eq8}) is simplified to $\langle \delta {V}_n^2
\rangle \sim 4\epsilon^2$, which is exactly two times the result
obtained by neglecting wall displacement. Consequently the root mean
square velocity as a function of the number of collisions is:
\begin{equation}
{V}_{rms,~HWA}(n) = \sqrt {4\epsilon^2\cdot n + \langle {V}_0^2
\rangle} \label{eq:eq9}
\end{equation}
The analytical result (\ref{eq:eq9}) based on the HWA is equally
shown in Fig.~1. The above map can also be  used to numerically
simulate the acceleration process of the particle, the corresponding
results being presented in Fig.~1. In contrast to the static wall
approximation which underestimates the acceleration of the
particles, the HWA provides an accurate description of this process,
indicating that the increased particle acceleration is due to the
dynamically induced correlation between the position and velocity of
the oscillating wall on collision. This hyper-acceleration can be
quantified by the ratio: $R_h\left(n\right)=\frac{\langle\delta
V_n^2\rangle_{Exact}}{\langle\delta V_n^2\rangle_{SWA}}=2$. However,
it should be noted that the specific value of $R_h$ depends on the
characteristics of the oscillation law and more specifically on its
turning points. For example, one can prove that for a piecewise
linear oscillation law $R_h$ is, in general, for any finite $n$
different than $2$ and only in the asymptotic limit becomes
$R_h\left(\infty\right) = 2$ \cite{Karlis:2006}.

\begin{figure}[htbp]
\centerline{\includegraphics[width=8.6 cm,height=6.45 cm]{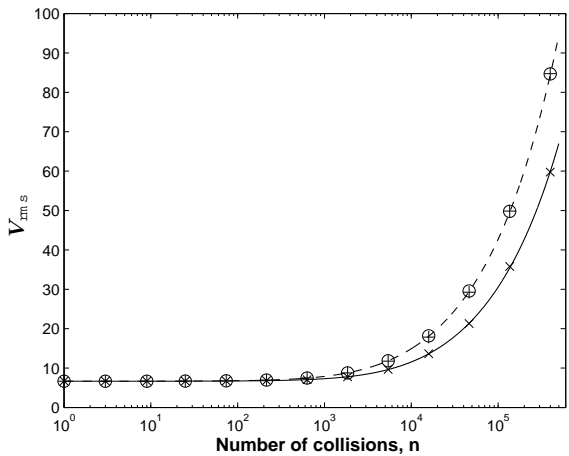}}
\caption{Numerical results for ${V}_{rms}$ of an ensemble of $10^4$
particles evolving in a FUM with two oscillating walls as a function
of the number of collisions. Results were obtained by iterating the
exact (circles) as well as the corresponding {\it static} (diagonal
crosses) and {\it hopping wall} (upright crosses) approximative
maps. It is noted that $V_{rms}$ is measured in units of $\omega w$.
Analytical results according to the SWA averaged over the random
phase (solid line) as well as the analytical prediction according to
eq.(\ref{eq:eq9}) (dash-dotted line) are also shown.}
\label{fig:fig1}
\end{figure}

The above analysis reveals the role of the fluctuations in the time
of flight $\delta {t}_n$ between successive collisions caused by the
displacement of the scatterer. Despite the existence of an external
stochastic component in the phase of the oscillating wall these
fluctuations lead to a {\it{systematic}} increase of the
acceleration of the particles. A simple explanation of the physical
mechanism leading to the increased acceleration becomes possible by
considering the various configurations of the collision processes
between the wall and the particles. Let us assume for a given
velocity of the incident particle that the wall is moving in the
same direction as the particle after passing the equilibrium
position. The collision time due to wall displacement increases
then, compared to the one assuming a wall fixed in space. In this
case, the velocity of the harmonically oscillating wall is a
decreasing function of time and therefore an increase of the
collision time leads to a decrease of the wall velocity on the
actual instant of the collision when compared to the {\it static
wall} approximation. This in turn leads to a lesser energy loss in
the course of the collision. This reasoning holds equally in all
other types of collision events, leading to the general picture of
less energy loss or more energy gain when the wall displacement is
taken into account.

The focus of our attention now shifts to the probability
distribution function (PDF) of the magnitude of the particle
velocity and number of collisions $n$, $\rho(|V|,n)$. It has been
shown that the change of $\langle {V}^2 \rangle$ as a function of
the number of collisions $n$ can be accurately described by a random
walk in momentum space $\langle V^2 \rangle \propto n$, provided
that the spatial motion of the walls is taken into account.
Numerical as well as analytical treatments in setups similar to the
present one suggest that $\rho(|{V}|,n)$ is described by a spreading
Gaussian \cite{Bouchet:2004,Ott:1993}. However, simulations with the
exact map of eq.(\ref{eq:eq1}) yield the histograms in Figs.~2a,b
$\rho(|V|,n)$ which show the PDFs corresponding to snapshots for $n
= 5 \cdot 10^4$, $5 \cdot 10^5$ collisions. It is shown that, even
if the initial PDF is a Gaussian, with increasing time, it is
transformed to a one-dimensional Maxwell-Boltzmann-like distribution
since the set of initial conditions leading to $|{V}| \ll 1$ is
vanishingly small for sufficiently long times. Using the {\it
hopping wall} map the following analytical expression for the PDF is
obtained, which describes the magnitude of the particle velocity for
$n \gg 1$:
\begin{equation}
\rho(|{V}|,n) = \frac{1}{\sigma^2} |{V}| e^{-\frac{{V}^2}{2
\sigma^2}} \label{eq:eq10}
\end{equation}
where $\sigma=\sqrt{\frac{4\epsilon^2 n + \langle
{V}_0^2\rangle}{2}}$. In Figs.~2a,b it is clearly seen that this
analytical result predicted by the HWA accurately reproduces the
exact behavior of the system. Furthermore, Fig.~2(c) shows the
evolution of the mean value $\langle \vert V \vert \rangle$ as a
function of the number of collisions $n$ obtained numerically using
the exact dynamics (open circles). For the sake of comparison we
also show the corresponding analytical result: $\langle \vert V
\vert \rangle = \sqrt{\frac{\pi}{2}} \sigma$ based on
eq.(\ref{eq:eq10}) (solid line).
\begin{figure}[htbp]
\centerline{\includegraphics[width=8.6 cm,height=6.45
cm]{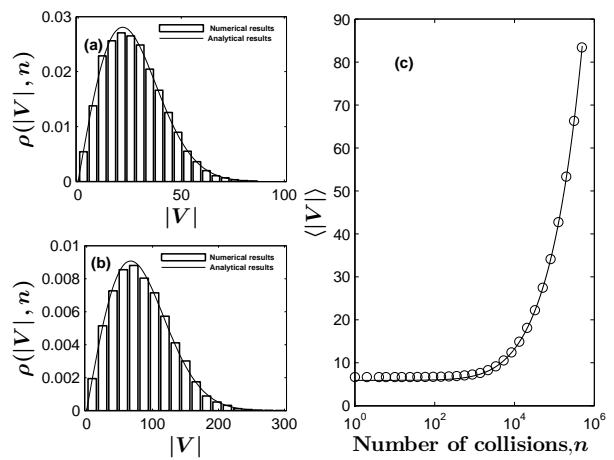}} \caption{Numerically computed PDF for the
magnitude of particle velocity using an ensemble of $10^4$
trajectories and following the exact dynamics of eq.(\ref{eq:eq1})
for: (a) $n=5 \cdot 10^4$ and (b) $n=5 \cdot 10^5$. In each case the
analytical result given by eq.(\ref{eq:eq10}) using the HWA is also
shown (solid line). Finally, (c) shows the numerically obtained
evolution of $\langle \vert {V} \vert \rangle$ as a function of $n$
using the exact dynamics (open circles) as well as the analytical
approximation based on eq.(\ref{eq:eq10}) (solid line).}
\label{fig:fig2}
\end{figure}

\begin{figure}[htbp]
\centerline{\includegraphics[width=8.6 cm,height=6.45 cm]{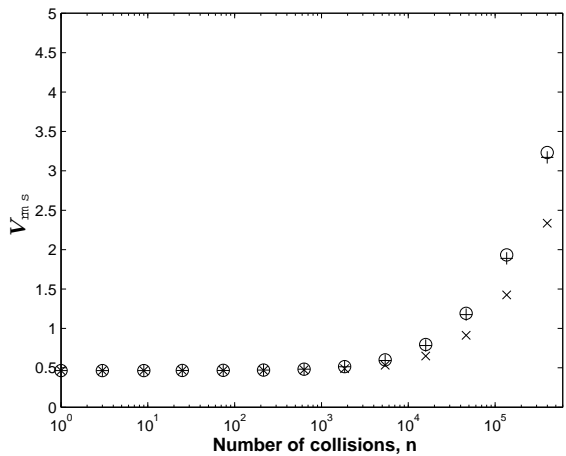}}
\caption{Numerical results for ${V}_{rms}$ of an ensemble of $10^4$
particles evolving in a triangular harmonically driven Lorentz gas
with randomly chosen direction of oscillation on each collision, as
a function of the number of collisions. Results were derived by
iterating the exact maps (circles) as well as the corresponding {\it
static} (diagonal crosses) and {\it hopping} (upright crosses)
approximations. The parameters $\left| {\rm {\bf A}} \right| =
0.01$, $\omega = 1$, $w=2.15$, $V_0 = \frac{1}{2.15}$ have been used
in the numerical simulations, with $A$ denoting the magnitude of the
amplitude of oscillation, $\omega$ the angular frequency, $w$ the
spacing between the disk centers at equilibrium and $V_0$ the
initial particle velocity. Velocities are measured in units $\omega
w$.} \label{fig:fig3}
\end{figure}

The development of hyper-acceleration takes place in
higher-dimensional scattering systems as well, such as a
time-dependent Lorentz gas consisting of harmonically oscillating
circular hard scatterers on a triangular lattice. It is emphasized
that in the time-dependent Lorentz gas system Fermi acceleration
exists without any externally imposed randomization
\cite{Loskutov,Leonel:2005}. However, the absence
 or presence
 of a random component in the dynamics influences the
acceleration law. For example, if the oscillation axis is fixed and
uniform throughout the lattice the acceleration law is $V_{rms}
\propto n^{1/4}$. On the other hand, if the oscillation axis  of the
disks is randomly chosen on each collision, simulating the effect of
thermal noise, the acceleration law becomes $V_{rms} \propto
n^{1/2}$, as in the 1D FUM system \cite{Karlis:2006}. In both cases,
the static approximation underestimates the ensemble mean energy
growth while the hopping approximation provides results much closer
to those of the exact model. To illustrate this, in
Fig.~\ref{fig:fig3} numerical results for the random setup outlined
above are presented. These are obtained utilizing the exact map
\cite{Papachristou:2001}, as well as the corresponding hopping and
static approximative maps \footnote{The derivation of the
approximative maps can easily be accomplished by: (a) setting the
amplitude of oscillation $A$ equal to $0$ in the equation describing
the spatial motion of the scatterer while retaining unchanged its
velocity amplitude (\textit{static approximation}) and (b) by
introducing a lag in time between the oscillation in configuration
and momentum space equal to the time of free flight between the
$(n-1)^{th}$ and $n^{th}$ collision --\textit{hopping
approximation}.}. Consequently, it can be inferred that the
development of hyper-acceleration is common to many driven dynamical
systems, and features in any billiard which allows Fermi
acceleration to develop. Moreover, the understanding gained in the
present investigation helps open up the prospect of designing
time-laws for the driving that provide a specified acceleration
behavior for the ensemble of particles thus leading to a desired
non-equilibrium i.e. time-evolving velocity distribution.

\end{document}